\documentclass[twocolumn]{aastex63}
\usepackage[version=4]{mhchem}
\usepackage{float}
\usepackage{parskip}
\received{}
\revised{}
\accepted{}
\shorttitle{Exploiting Network Topology for Inference of Surface Reaction Networks}
\shortauthors{Heyl et al.}
\graphicspath{{./}{figures/}}

\begin{document}

\title{Exploiting Network Topology for Accelerated Bayesian Inference of Grain Surface Reaction Networks }

\correspondingauthor{Johannes Heyl}
\email{johannes.heyl.19@ucl.ac.uk}

\author[0000-0003-0567-8796]{Johannes Heyl}
\affiliation{Department of Physics and Astronomy, University College London, Gower Street, WC1E 6BT, London, UK}

\author[0000-0001-8504-8844]{Serena Viti}
\affiliation{Leiden Observatory, Leiden University, PO Box 9513, 2300 RA Leiden, The Netherlands}
\affiliation{Department of Physics and Astronomy, University College London, Gower Street, WC1E 6BT, London, UK}

\author[0000-0003-4025-1552]{Jonathan Holdship}
\affiliation{Leiden Observatory, Leiden University, PO Box 9513, 2300 RA Leiden, The Netherlands}
\affiliation{Department of Physics and Astronomy, University College London, Gower Street, WC1E 6BT, London, UK}

\author[0000-0003-2268-2519]{Stephen M. Feeney}
\affiliation{Department of Physics and Astronomy, University College London, Gower Street, WC1E 6BT, London, UK}



\begin{abstract}

In the study of grain-surface chemistry in the interstellar medium, there exists much uncertainty regarding the reaction mechanisms with few constraints on the abundances of grain-surface molecules. Bayesian inference can be performed to determine the likely reaction rates. In this work, we consider methods for reducing the computational expense of performing Bayesian inference on a reaction network by looking at the geometry of the network. Two methods of exploiting the topology of the reaction network are presented. One involves reducing a reaction network to just the reaction chains with constraints on them. After this, new constraints are added to the reaction network and it is shown that one can separate this new reaction network into sub-networks. The fact that networks can be separated into sub-networks is particularly important for the reaction networks of interstellar complex organic molecules, whose surface reaction networks may have hundreds of reactions. Both methods allow the maximum-posterior reaction rate to be recovered with minimal bias. 

\end{abstract}



\section{Introduction} \label{sec:intro}
Interstellar dust is a crucial part of the interstellar medium. The dust grains are responsible for much of the rich chemistry observed, as they act as an energy sink, allowing reactions between adsorbed atoms and molecules \citep{Dishoeck_dust}. There is strong evidence to suggest that complex-organic molecules form on these interstellar dust grains \citep{Herbst_van_Dishoeck}. In fact, this rich chemistry is thought to take place long before stars have formed, during the dark cloud phase. In the last couple of decades many experiments have been performed to determine the surface reactions  occurring on the icy mantles (e.g. see review by \cite{Linnartz}). However, surface reactions in the laboratory are typically only 
investigated within a narrow range of physical parameters constrained by what is available with the experimental techniques employed 
and thus often not fully representative of the ISM conditions. Hence the available experimental data for interstellar ices are limited.

Bayesian inference has become a standard tool in astrophysics for determining model parameters from observations. In recent years, it has also become a tool in astrochemistry \citep{holdship, Antonios, Damien}. However, by considering increasingly rich chemistry one must ultimately consider more complicated reaction networks. This results in an increased computational cost. There exists much in the literature regarding chemical network reduction as a means of reducing the computational complexity of the problem being solved \citep{tennyson}. An understanding of chemical reaction networks and how to simplify them has become increasingly crucial in astrochemistry \citep{Xu_2019, Grassi}. However, these methods have primarily focused on simplifying the network for the forward problem. For example, \cite{Xu_2019} adopted an iterative approach, where they evaluated the importance of each species at each timestep. The advantage of Bayesian inference is that it provides probability distributions for the parameters of interest conditioned on the available data, thereby allowing us to quantify the uncertainties on these parameters. However, there is the issue that not all the parameters of interest can be determined. The iterative approach mentioned above only focuses on the most reactions to which species are the most sensitive. This allows for the reduction in computational expense, unlike for Bayesian inference. In this work we look to use various features of the network topology to reduce the computational expense of the inference process.


In this work, we build upon the work done by \cite{holdship} (hereafter H18) and use the same reaction network to highlight how aspects of the geometry of the network can be exploited to determine the model parameters at reduced computational expense. It should be noted that this reaction network only has 24 reactions with four constraints, but this technique should generalise to larger networks. It is hoped this will prove particularly useful when considering reaction networks of complex organic molecules (COMs), where the number of reactions is large and the number of constraints small.

We begin by first presenting the chemical network used in Section 2.  Following this, we argue why we can reduce the network with its constraints to a simpler one as well as present explanations for how the positions of particular constraints in the networks are crucial. We present the results of the Bayesian inference for this. We then go on to discuss how specific aspects of the topology of the reaction network are useful to consider and how these influence our choices when we reduce the network. Finally, we look at how we can separate a reaction network into smaller sub-networks.


\section{The Chemical Network}
\begin{figure}[h]
\centering
\includegraphics[width=0.5\textwidth]{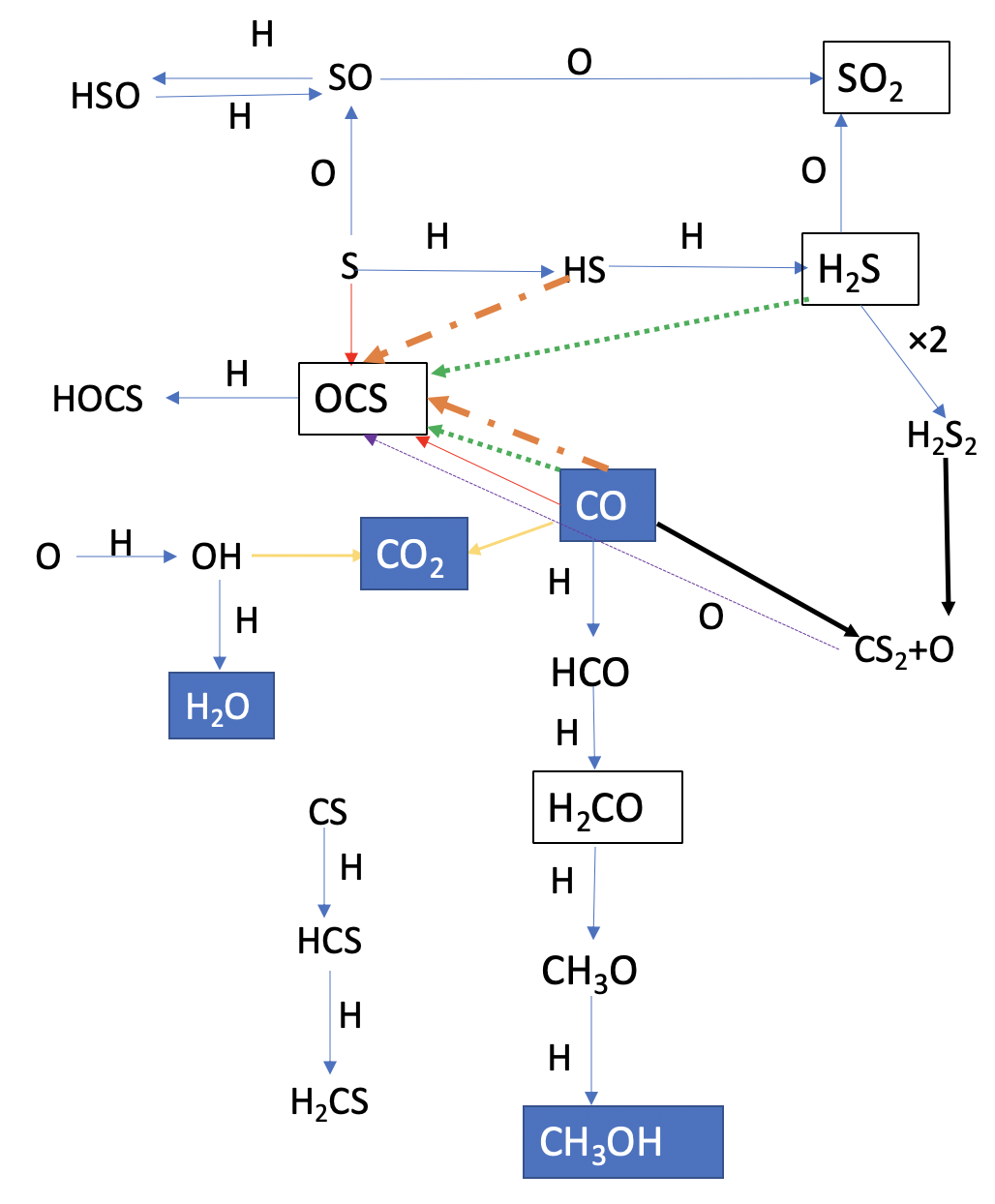}
\caption{A diagram of the chemical reaction network considered. For the sake of simplicity, any reactions with hydrogen and oxygen are represented with H and O next to the arrow. For the case where a molecule can be formed in multiple reactions, such as for OCS, the arrow colours pointing to that molecule indicate the reactants. For example, the dash-dotted orange arrows that point from HS and CO to OCS indicate that these two molecules form OCS. Molecules in blue boxes have constraints on their final abundances. Molecules in white boxes have upper limits on their abundances.}
\label{reaction_network}
\end{figure}

We use the same chemical network as in H18, which we set out pictorially in Figure \ref{reaction_network}. We list all reactions in Table \ref{reaction_network_table}, assigning them numbers which we use throughout the paper for brevity. For the sake of simplicity, the hydrogen abundance is not a conserved quantity, as it is typically $\sim10^{4}$ times more abundant than any other molecule, so its loss in this reaction network is negligible. We also emphasise that this network is a toy model and is not meant to properly reflect a complete grain surface network but simply serves as a proof-of-concept. 

A modified version of the open source gas-grain code UCLCHEM was used \citep{UCLCHEM_release_paper}. This version only considered the surface chemistry of a collapsing dark cloud. The cloud is modelled as collapsing from a density of $10^{2}$ cm$^{-3}$ to 2 $\times$ 10$^{4}$ cm$^{-3}$ over 10 million years. The collapse takes place isothermally at 10 K. As chemistry only occurs on the grain surface, freeze-out rates are required. Freeze-out rates are the rates at which atomic and molecular species stick to the dust grains and build up ices \citep{Hocuk, Fraser}. The freeze-out rates were found in H18 by running a single-point model of the full UCLCHEM code. The species that were given freeze-out rates were: CO, CS, O, H, OH and S. More details can be found in H18. 

As already mentioned above, the small chemical network we use for this work is not meant to represent a comprehensive surface network. Nevertheless,  the choice of most of the reactions was based on the results of experimental studies. For example, the successive hydrogenation of CO to form CH$_{3}$OH has been studied in detail and is considered to be the dominant reaction pathway \citep{Chuang}. Similarly, CO$_{2}$ has been found to be efficiently formed when CO and OH react \citep{Ioppolo}. Beside a small network representing the main routes of carbon- and oxygen-bearing species on the ices, we chose to include a small network producing sulphur-bearing species, since there is still much unknown about the form that ultimately sulphur takes on the ices during the cold phase of the star formation process \citep{Woods, sulphur_paper}.

In this work we consider a number of variants of the chemical network shown in Figure 1. These configurations, which differ in terms of the reactions and/or constraints used, are enumerated in Table \ref{reaction_configuration_table}: the configuration numbers will be used throughout the work. The combination of the full reaction network shown in Figure \ref{reaction_network} with the abundance measurements as listed in Table \ref{first_abundance_table1} is Configuration 1.

\begin{table}
 \begin{tabular}{||c c||} 
 \hline
 Reaction No. & Reaction  \\ [0.5ex] 
 \hline
 1 & \ce{O + H -> OH}  \\ 
 \hline
 2 & \ce{O + H -> H_{2}O}   \\
 \hline
 3 & \ce{CO + OH -> CO_{2}} \\
 \hline
 4 & \ce{S + H -> HS}  \\
 \hline
 5 & \ce{HS + H -> H_{2}S}  \\  
 \hline
  6 & \ce{H_{2}S + S -> H_{2}S_{2}}   \\  
   \hline
  7 & \ce{CS + H -> HCS}   \\  
 \hline
  8 & \ce{HCS + H -> H_{2}CS}   \\  
 \hline
  9 & \ce{CO + S -> OCS}   \\  
 \hline
  10 & \ce{OCS + H -> HOCS}   \\ 
 \hline
  11 & \ce{H_{2}S + CO -> OCS}   \\ 
 \hline
  12 & \ce{H_{2}S + H_{2}S -> H_{2}S_{2}}   \\  
 \hline
  13 & \ce{H_{2}S_{2} + CO -> CS_{2} + O}   \\ 
   \hline
  14 & \ce{H_{2}S + O -> SO_{2}}   \\ 
 \hline
  15 & \ce{CS_{2} + O -> OCS + S}   \\ 
 \hline
  16 & \ce{CO + HS -> OCS}   \\ 
 \hline
  17 & \ce{S + O -> SO}   \\ 
 \hline
  18 & \ce{SO + O -> SO_{2}}   \\ 
 \hline
   19 & \ce{SO + H -> HSO}   \\  
 \hline
  20 & \ce{HSO + H -> SO}   \\ 
 \hline
  21 & \ce{CO + H -> HCO}   \\  
 \hline
  22 & \ce{HCO + H -> H_{2}CO}   \\  
 \hline
  23 & \ce{H_{2}CO + H -> H_{3}CO}   \\  
 \hline
   24 & \ce{H_{3}CO + H -> CH_{3}OH}   \\  
 \hline
\end{tabular}
\caption{Table of the reactions used in this work taken from \citep{holdship}}
\label{reaction_network_table}
\end{table}

\section{Bayesian Inference}

\begin{table*}[!htbp]
\begin{centering}
\begin{tabular}{|l|l|l|}
\hline
Configuration No. & Reactions Used & Molecules with Constraints \\ \hline
              1            & 1-24                &  CO, CH$_{3}$OH, CO$_{2}$, H$_{2}$O                \\ \hline
              2           &  1-6, 9-24              & CO, CH$_{3}$OH, CO$_{2}$, H$_{2}$O                 \\ \hline
              3           &   1-3, 21-24             &  CO, CH$_{3}$OH, CO$_{2}$, H$_{2}$O                \\ \hline
              4            &  1-24              &    CH$_{3}$OH, CO$_{2}$, H$_{2}$O              \\ \hline
              5            &     1-24           &     CO,  CO$_{2}$, H$_{2}$O              \\ \hline
              6            &     1-24           &     CO, CH$_{3}$OH, CO$_{2}$, H$_{2}$O, SO$_{2}$, OCS, H$_{2}$S             \\ \hline
              7            &     4-20           &     CO, SO$_{2}$, OCS, H$_{2}$S             \\ \hline
              8            &     4-20           &     SO$_{2}$, OCS, H$_{2}$S             \\ \hline
              9            &     1-3, 21-24           &     CH$_{3}$OH, CO$_{2}$, H$_{2}$O              \\ \hline

\end{tabular}
\caption{A table listing all the various network configurations used and referred to throughout this work.}
\label{reaction_configuration_table}
\end{centering}
\end{table*}

\subsection{Introduction to Bayesian Inference}
As in H18, our aim is to determine the 24 reaction rates, which we represent as a vector,  $\textbf{k} = (k_{1}, {k_{2}} ... k_{24}$). We are therefore faced with a 24-dimensional inference problem. The grain code used took in these reaction rates and produced the corresponding abundances, which are represented by a vector $\textbf{Y} = (Y_{1}, {Y_{2}} ... Y_{23}$). Henceforth we refer to the ``forward model" when we input a particular value of $\textbf{k}$ to obtain some $\textbf{Y}$. 

We know the abundances of four of the molecules in the ices of the network shown in Figure \ref{reaction_network}, which are in blue boxes. These are taken from \cite{Boogert} and listed in Table \ref{first_abundance_table1}. We are looking to solve the ``inverse problem", i.e. what values of $\textbf{k}$ yield values of $\textbf{Y}$ that match the observations best? Such a problem naturally lends itself to a Bayesian approach. The inherent degeneracy of this problem should be noted. Observations only exist for 4 of the 24 molecules. This suggests that the rates of the reactions that do not influence the abundances of these four molecules will be poorly constrained (if at all), and there will be many values of these rates that give the same observations. For a discussion of the degeneracies, please refer to H18. Exploiting the low number of constraints in order to speed up the inference process is a crucial point in this work.

We use Bayes' Law to determine the probability distribution of the values of the reaction rates 
\begin{equation}
P(\textbf{k} \vert \textbf{d}) = \frac{P(\textbf{d} \vert \textbf{k})P(\textbf{k})}{P(\textbf{d})},
\end{equation}
where $P(\textbf{k} \vert \textbf{d})$ is the posterior probability distribution, $P(\textbf{k})$ is the prior, $P(\textbf{d} \vert \textbf{k})$ is the likelihood and $P(\textbf{d})$ is referred to as the evidence. Here $\textbf{d}$ is the data, i.e. the observed values of the abundances.  The evidence is a normalising factor and is typically difficult to evaluate. However, as it is independent of $\textbf{k}$, we can instead just consider
\begin{equation}
P(\textbf{k} \vert \textbf{d}) \propto P(\textbf{d} \vert \textbf{k})P(\textbf{k}),
\end{equation}
which is just the unnormalised posterior.

\subsection{Implementation}
Evaluating the reaction rate posterior requires specification of a prior on the reaction rates and a likelihood. We chose the same prior as in H18: a log-uniform distribution between $10^{-30}$ and $10^{-5}$. From chemical considerations, we know that these reaction rates are ultimately very fast. Typically, we might therefore select a prior that favours higher reaction rates. We chose, however, to ignore this information, instead following the traditional approach of using a log-uniform prior, which equally weights rates over a range of orders of magnitude.\footnote{It is worth noting that the use of a single prior to represent complete ignorance \citep{Walley, Norton} has received criticism. Complete ignorance can be represented by repeating the analysis using several priors that significantly differ from one another (see \cite{Fischer} for a straightforward application to a simple problem of chemical kinetics).} It is important to note that, despite our motivation for this prior, we realise that we are in a prior-dominated regime, which we demonstrate in Appendix B with very different prior assumptions. Unlike in the data-dominated regime, these prior-dominated posteriors differ significantly among themselves. This, however, does not detract from the analysis we conduct in this work.


A Gaussian likelihood function was used, which takes the form 

\begin{equation}\label{likelihood}
\centering
P(\textbf{d} \vert \textbf{k}) =  \prod_{i=1}^{n_{d}} \frac{1}{\sqrt{2\pi}\sigma_{i}} \exp\left({-\frac{(d_{i}-Y_{i})^{2}}{2\sigma_{i}^{2}}}\right),
\end{equation}

where $n_{d}$ is the number of observations and $\sigma_{i}$ is the uncertainty of the $i$th observation. We only multiply over the species which have observed abundances. We refer to these observed abundances as constraints as they constrain the parameter space of our reaction rate posteriors.

In order to determine the posterior, the PyMC3 Python package was utilised \citep{pymc3}. The PyMC3 package includes a range of samplers. Here, we used the Metropolis-Hastings algorithm, a simple Markov Chain Monte Carlo (MCMC) method. A Gaussian proposal distribution was used. Before each run, 500 tuning steps were initially taken to determine the optimal covariance of the proposal distribution. These tuning steps were not included in our analysis and were discarded. For the relatively low number of dimensions ($<50$), a point-based sampler such as this one is suitable. 50 chains of length $10^{6}$ were used to sample the posterior probability space. We created a Python wrapper of the grain code using F2Py that was then fed values of \textbf{k} during the sampling process \citep{F2Py}.

Though simple sampling methods suffice for the dimensionality of the posteriors considered here, the same will not be true for more complex networks. A key point that needs to be considered is the limitations of many MCMC methods as the number of dimensions increases. \cite{DNest4} discuss how some widely-used samplers struggle to give sensible results in higher dimensions. The popular emcee Python package that was used in H18 is discussed to be useful for when the number of dimensions is fewer than 50, with it struggling in higher dimensions \citep{emcee_paper}. Even for the case where the sampler does not struggle as the number of dimensions increases, the time taken to run the inference process will still increase. This increase in computation time will eventually become prohibitive. In section \ref{sulphur_application} we will argue that we can split our reaction network up into sub-networks on which we can perform Bayesian inference. By carefully placing the ``cut" on the reaction network, we then show that we can reproduce the results of the full reaction network inference with these sub-networks. Each sub-network has lower dimensionality than the original network, meaning its rates can be inferred more quickly and by simpler samplers. Additionally, the inference process on all the sub-networks can be run in parallel.
\subsection{Constraints}\label{constraints}

\begin{table}[ht]
\centering
 \begin{tabular}{||c c||} 
 \hline
 Species & Abundances relative to H  \\ [1ex] 
 \hline\hline
 H$_{2}$O & $(4.0 \pm 1.3) \times 10^{-5}$  \\ 
 \hline
 CO & $(1.2 \pm 0.8) \times 10^{-5}$\\
 \hline
 CO$_{2}$ & $(1.3 \pm 0.7) \times 10^{-5}$  \\
 \hline
 CH$_{3}$OH & $(5.2 \pm 2.4) \times 10^{-6}$  \\
 \hline
\end{tabular}
\caption{The abundances and uncertainties for the molecules with observed values taken from \cite{Boogert}.}
\label{first_abundance_table1}
\end{table}

It is essential to include constraints in order to formulate a likelihood function. In H18, four constraints for molecules in the reaction network were taken. Table \ref{first_abundance_table1} shows the abundances of the molecules taken from \cite{Boogert}. These ice abundances are derived from ice band profiles. The column densities can be calculated using the integrated optical depth as well as the integrated band strength. The latter were determined from laboratory experiments by \cite{Boogert}. From the table, it is clear that the strongest constraint is on H$_{2}$O, which is known to exist at the $3\sigma$ level, whereas the other molecules' abundances differ from zero at only $1.5-2.2\sigma$. 
H18 also reformulated the likelihood function to include upper bounds on the abundances of OCS, H$_{2}$S, SO$_{2}$ and H$_{2}$CO. This was not found to have a significant effect on the reaction rates determined and so we do not include these upper bounds in the following work. For the rest of this work we use a likelihood function of the form in Equation \ref{likelihood}.

\section{Network Reduction Methods}

\begin{figure*}[h]
    \includegraphics[width=1\textwidth]{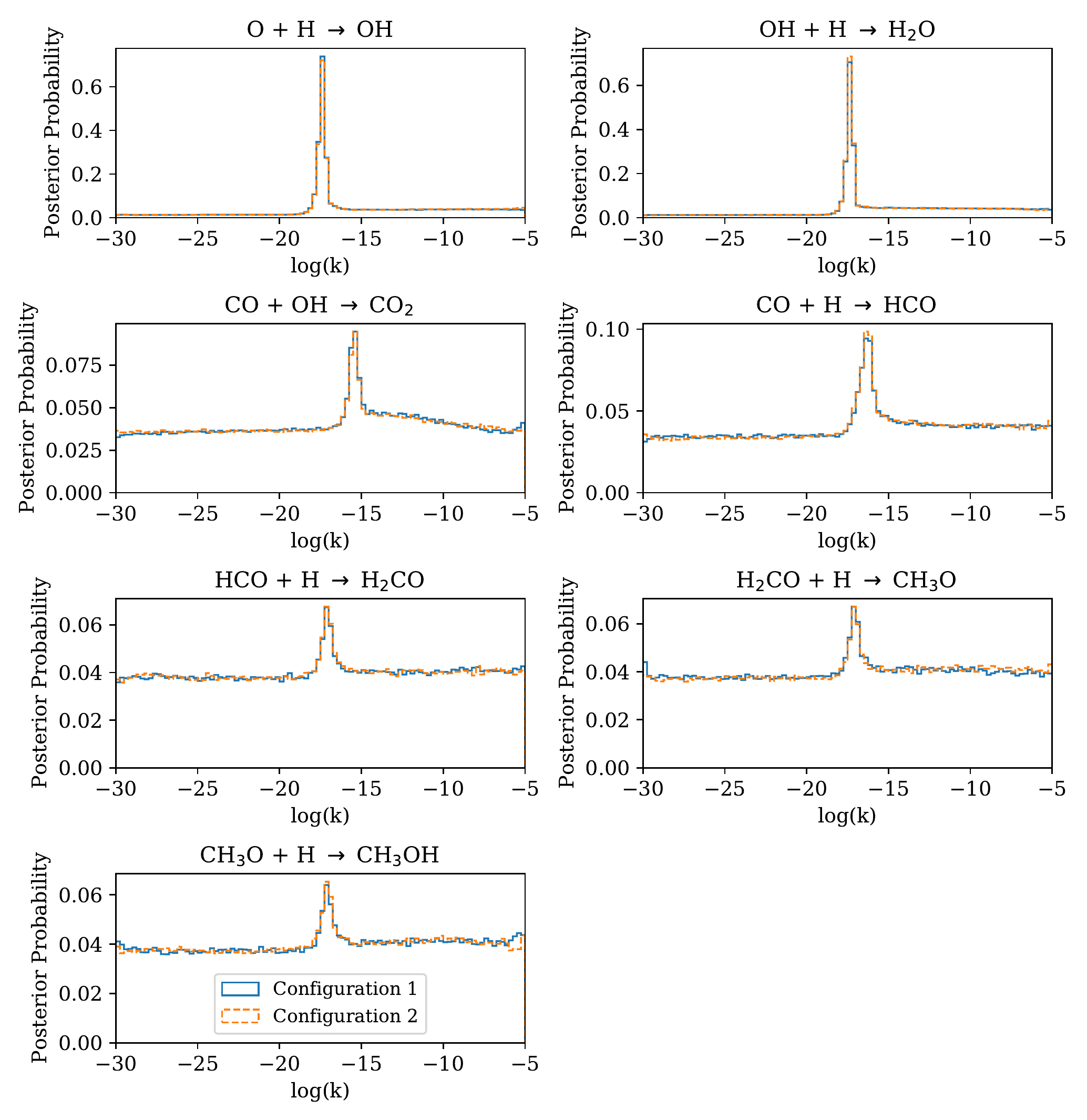}
    \caption{Plots of the posterior probability distribution for the original reaction network considered in \cite{holdship} as well as the 22-dimensional effective network by removing the ``H$_{2}$CS chain". We observe good agreement in the shapes of the posterior distributions, with any differences due to specific samples drawn from the MCMC chains. The configuration 1 posteriors match those from H18.}
    \label{Effective network vs Holdship}
\end{figure*}

\subsection{Overview}
\cite{Galagali} considered a similar problem to the one discussed in this work in the context of systems biology. There, they considered the case of a reaction network with a single observation. The main differences between the networks they considered and the one being considered here is the absence of enzymes as well as the absence of reversible reactions. 

They defined the ``effective reaction network" as the subset of reactions that must be kept in order to produce the same values of the observable. This is a useful concept to consider, especially in the context of network connectivity. Some subsets of reactions will evolve completely independently of one another, with there being no competition for chemical species. While this may seem unlikely in the context considered here, it is true when one assumes that hydrogen's abundance is significantly higher than that of any other species. This can be seen in figure \ref{reaction_network}, where the successive hydrogenation of CS to form H$_{2}$CS is clearly independent of the rest of the reaction network. This reaction chain will be referred to as the ``H$_{2}$CS chain" from now on.

\subsection{Network Reduction of Non-Connected Networks}\label{effective_network}
We consider the ``H$_{2}$CS chain" in greater depth. Under the assumption that there is no competition for hydrogen, we should expect the ``H$_{2}$CS chain" and the other reaction network with the remaining 22 reactions (Configuration 2) to evolve completely independently, as the species do not interact. We infer the reaction rate posteriors for  Configurations 1 and 2. Figure \ref{Effective network vs Holdship} shows the posteriors on the common reactions to be essentially identical. Any differences are due to the specific samples drawn in the MCMC chains. This intuitively makes sense as the ordinary differential equations that govern the two sub-networks will evolve independently of one another. Since none of the reactions in the ``H$_{2}$CS chain" are constrained in Configuration 1, it stands to reason that these reaction rates are nuisance parameters. As such, removing these two reactions from the inference should make no difference. 

While the example given might seem trivial, one needs to consider under what circumstances one might have a reaction network with a disconnected segment. In surface grain chemistry, the molecules must contend with both an activation energy barrier as well as a diffusion energy barrier. Only if both of these can be overcome, is the reaction likely to happen efficiently. If either one of these barriers is too high, then one can in fact approximate that reaction as not happening and ``cut" off that reaction. This might lead one to separating a reaction chain from the rest of the network. In this example, one could conceivably imagine a hypothetical reaction being possible between H$_{2}$CS and any other molecule in Configuration 2, but the activation energy barrier is too high to overcome at 10 K. It is therefore simpler to exclude it.


\section{Further Network Reduction}

In the previous section, we observed that the network connectivity of a chemical reaction network can allow us to discard reactions that do not influence the values of observed abundances. In this section, we develop this idea further by arguing that the locations of the constraints in the reaction network allow us to discard more reactions. 

We wish to emphasise once again that we are not seeking to make quantitative predictions about the reaction rates. Instead, we are looking to develop a qualitative understanding of the kinetics as well as develop an intuition for how the methods that will be discussed in this work can be applied to other astrochemical modelling scenarios.

\subsection{Reducing the Network}\label{network_reduction}

We begin by briefly returning to the posteriors in Figure \ref{Effective network vs Holdship}. The uncertainties on the rates of reactions 1 and 2 are significantly smaller than for reactions 3 and 21-24. This relates to the uncertainties on the abundances of the molecules, as discussed in Section \ref{constraints}. In the limit that the abundances of all molecules involved in a reaction are perfectly known, one would expect the posterior distribution of the reaction rate to approach a Dirac delta function. The greater confidence level in water's presence is therefore responsible for the tighter constraints on the rates of Reactions 1 and 2. Improving the precision of the abundance measurements of CO, CO$_{2}$ and CH$_{3}$OH would in turn tighten the constraints on their reaction rates. 
In Figure \ref{reaction_network}, we observe that all the reactions whose reaction rates are constrained have a constraint at the end of their respective chain. Consider the successive hydrogenation of carbon monoxide to form methanol, henceforth referred to as the ``methanol chain". The fact that there is a constraint present at the end is significant. By constraining the amount of methanol, one effectively constrains the reaction rates of its precursors, CH$_{3}$O and H$_{2}$CO from below, as the existence of methanol requires its precursors to have been produced. If the reaction rate of these reactions is too high, then too much methanol will be produced. However, there is an inherent degeneracy in the rates of the intermediate reactions, so it is unclear how the rates are partitioned between the two reactions. What one finds is that these intermediate reaction rates are coupled, by observing their joint probability distributions. One reaction will serve as the rate-limiting reaction, with the other compensating for this by being significantly faster to produce a sufficient amount of the final product, in this case methanol. This is discussed in more depth for the specific reactions in H18.

Additionally, the constraint on carbon monoxide constrains the abundances of these molecules from above. One can then qualitatively say that as there are now constraints on their abundances, there are therefore constraints on their reaction rates, as the reaction rates reflect how much of these molecules forms over time. 

\begin{figure*}[htp]
\includegraphics[width=\textwidth]{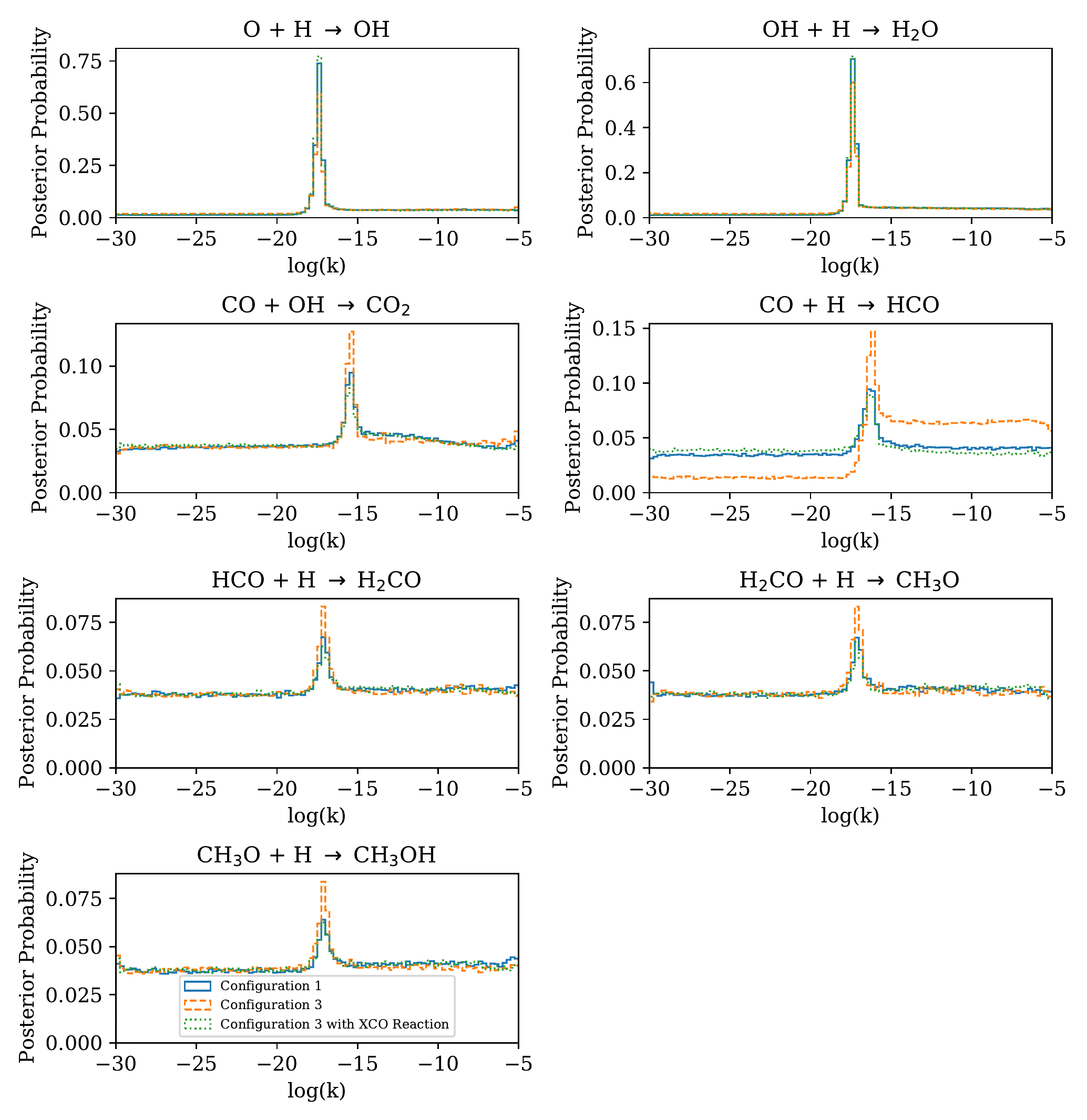}
\caption{Plots of the posteriors of Configuration 1, Configuration 3 and Configuration 3 with the dummy reaction $X+CO \rightarrow XCO$. We observe that the inclusion of this additional dummy reaction provides a better approximation to the Configuration 1 posterior than the Configuration 3 posterior does. }
\label{Sulphur_freeze_out_rate}
\end{figure*}

In the network shown in Figure \ref{reaction_network}, none of the sulphur-bearing molecules have constraints on them. The reaction rates of these reactions can be treated as unconstrained parameters, despite the fact that carbon monoxide, a central molecule in the sulphur-centric network, is constrained. This suggests that the sub-network consisting of the 10 non-sulphur bearing species can be treated independently. We exclude these sulphur-bearing reactions for now and only consider reactions 1-3 and 21-24. This is Configuration 3, which is a ``sub-network" of Configuration 1 as it contains a subset of the reactions. 

To investigate the impact of excluding the sulphur-bearing reactions, we re-run our Bayesian inference on this configuration, producing the posterior probability distribution functions in Figure \ref{Sulphur_freeze_out_rate}. With the exception of $H+CO \rightarrow HCO$, we recover the maximum-posterior reaction rates obtained previously. However, we see that the variances have decreased for reactions 3 and 21-24, resulting in more peaked distributions, and reactions 1 and 2 have posteriors that are less peaked than for Configuration 1. 

We would like to emphasise that this is a purely artificial effect. By eliminating the sulphur-bearing reaction rates, which were essentially nuisance parameters, the variance of reactions 3 and 21-24 have decreased. It should be noted that these reactions compete for CO with the removed sulphur sub-network. Removing the sulphur-based reactions means the non-sulphur reactions have to use up more CO. We observe that of all the reactions, the reaction $H+CO \rightarrow HCO$ sees the greatest change between Configuration 1 and Configuration 3. In fact, we observe that a significant portion of the posterior mass is shifted from the reaction rates below the peak to the reaction rates above the peak. This, coupled with the increase in the maximum-posterior rate, suggests that the excess CO, that would normally be consumed by the sulphur reactions, is stored in the methanol chain. In particular, the fact that only the hydrogenation of CO experiences a significant change in the posterior suggest that the excess CO is stored as HCO. 

At this point, it is unclear why Reactions 1 and 2 see increases in the variance of their posteriors. For these reactions, the decrease in the posterior mass under the peaks is compensated for by an increase in the posterior masses for reaction rates slower than the maximum posterior-rate. This suggests that the reactions can proceed at slower rates and still produce sufficient OH that goes on to produce H$_{2}$O and CO$_{2}$.  

The MCMC runs for Configuration 3 are 2.3 times shorter than for Configuration 1, with the time taken for the runs decreasing from 30 to 13 hours. By excluding the unconstrained reactions, we are able to reduce walltime drastically, at the cost of moderate changes to the posterior. In the next sub-section, we discuss a method for recovering the full posterior. Finally, it should be emphasised that this dimension reduction must only be considered when solving the inverse problem. For a full picture of the chemistry one must include all reactions in the forward-model. 

\subsection{Recovering the Full Variance}\label{variance_recovery}
We noted previously that the decrease in the variance of the posteriors was an artificial effect. Recovering as much of this original variance as possible is critical, as without it the precision of the inferred rates will be overstated. A consideration of the reactions that were removed is a good starting point. Looking at Figure \ref{reaction_network}, one can see that from the perspective of Configuration 3, the removal of the reactions affected the CO depletion. In other words, Configuration 3 only sees that we removed CO-depleting reactions. The products of these CO-depleting reactions are not constrained, and these reactions are therefore responsible for the larger uncertainty in the posteriors of Configuration 1 compared to Configuration 3.

In order to recover this variance, one must account for the artificially altered CO depletion. As a first, simple approximation we add a fake reaction $X+CO \rightarrow XCO$, where X is meant to encompass all the removed reactions that consumed CO, and XCO is simply the network of products. The abundances of both X and XCO remain unconstrained. Even though X is not meant to represent a particular molecule, it still requires a freeze-out rate, as this grain-code only considers reactions that take place on the grain surface. The freeze-out rate for sulphur was used. The posteriors are shown in Figure \ref{Sulphur_freeze_out_rate}. Adding a single unconstrained reaction clearly yields a good (though still imperfect) approximation to the full set of sulphur-consuming reactions in this setting, matching the variance of the full network's rate constants more closely and removing the bias on the inferred rate of $H+CO \rightarrow HCO$. Further work should be done to investigate whether increasing the number (and architecture) of ``dummy reactions” aids in recovering the full posterior.

\subsection{Network Topology Considerations} \label{experimenting_with_linear_topology}
We now discuss how the placement of the constraints in the network can be significant. From the above analysis, it is not entirely clear what constraints are the most ``essential". To shed light on this problem, we focus on the methanol chain, which has a constraint on molecules at both ends. 

We perform Bayesian inference on the full reaction network twice:
\begin{itemize}
    \item once without the CO constraint (Configuration 4)
    \item once without the CH$_{3}$OH constraint (Configuration 5)
\end{itemize}

\begin{figure*}[htp]
\includegraphics[width=\linewidth]{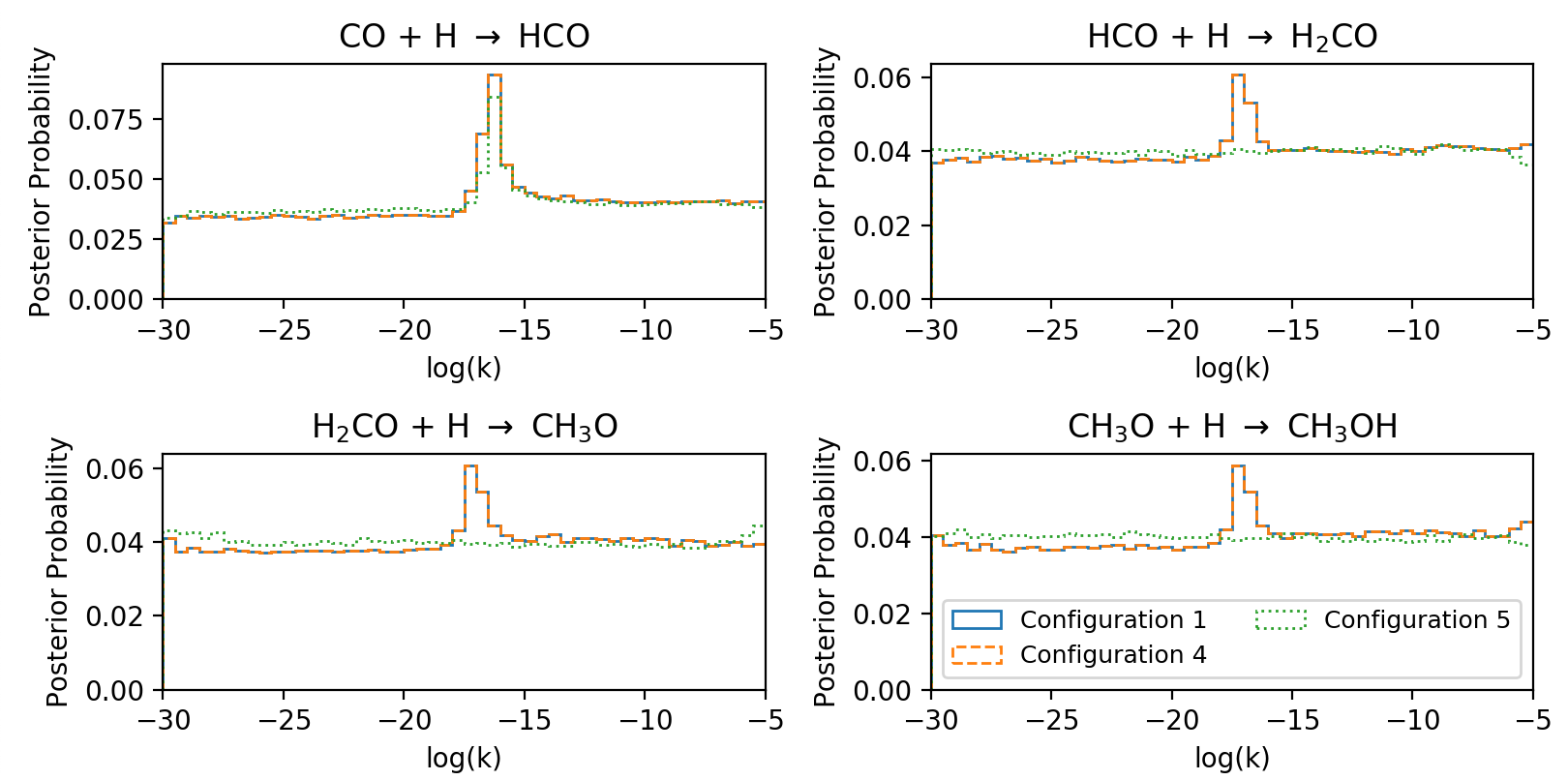}
\medskip
\caption{The posterior probability distributions for reactions 21-24 when CO and methanol are separately removed. The original distributions are also included for comparison. We observe that for reactions 21-24, removing CO neither affects the position of the peak of the distribution nor the shape of the distribution. Removing methanol does not change reaction 21's maximum-posterior rate, but removes all information about the reaction rates of reactions 22-24. We do not include reactions 1-3, as their posteriors are unchanged when the constraints are removed.}
\label{Removing CO and Methanol}
\end{figure*}

The resulting posterior probability distribution functions for Reactions 21-24 are shown in Figure \ref{Removing CO and Methanol} alongside the distributions for the original network. The posteriors for Reactions 1-3 are not included, because these did not change significantly.

We observe that removing the constraint on CO has no effect on the reaction rates. The reaction rate posteriors for reactions 21-24 are broadly identical. This can be easily explained. In this case, one knows that a fixed amount of methanol is produced. As such, over the period of 10 Myr, a certain amount of CO has to be consumed. This results in the successive hydrogenations being constrained, which is why the reaction rate posteriors do not change.

However, removing the constraint on methanol results in the loss of constraints on 3 of the 4 reaction rates of the methanol chain, with only the hydrogenation of CO being recovered. This appears to suggest that there is a notion of ``distance" between a constraint and the reaction rate of interest. Information about the subsequent reaction rates in the methanol chain is lost due to methanol's abundance being unconstrained. In this model, we know that CO (as one of the adsorbed species) is present on the grains, but there is no information about how much of it goes into making methanol. However, it is interesting that reaction 21 remains constrained. One possible interpretation is that we know how much CO is used in reaction 3 and this possibly helps constrain how much CO is used in reaction 21. However, using this reasoning, one cannot explain why reactions 9 and 13 are not constrained, as these are also CO depletion reactions. 

This suggests that in any reaction chain, some knowledge of the abundance of the end-products is required, which might be problematic when the species are undetected. However, one can still provide theoretical predictions for these abundances that could be used.

\section{Application to the Network with Artificial Sulphur Constraints}\label{sulphur_application}
As a proof of concept, we wish to apply the insight from the previous section to a new grain surface network. However, this is difficult due to the limited number of observations that exist for grain-surface molecules. \cite{Boogert} provided upper limits for several molecules in the ice. We chose to artificially transform the upper limits on OCS, H$_{2}$S and SO$_{2}$, into weak measurements by taking their abundances to be half the respective upper limit with an uncertainty of one-quarter of the upper limit. We would like to emphasise again here that we are simply trying to demonstrate how the location of these three additional constraints provides us with more knowledge of $\textbf{k}$. We do not claim this to be an accurate representation of sulphur chemistry on the ices in a dark cloud. Many theoretical and modelling studies have been recently performed investigating the sulphur depletion and the reactions on surfaces involving sulphur-bearing species and we refer the reader to such studies for a comprehensive review on the subject (e.g. \cite{Jimenez, Woods,Vidal_Wakelam,sulphur_paper}).

\begin{table}\label{first_abundance_table}
 \begin{tabular}{||c c||} 
 \hline
 Species & Abundances  \\ [1ex] 
 \hline\hline
 H$_{2}$O & $(4.0 \pm 1.3) \times 10^{-5}$  \\ 
 \hline
 CO & $(1.2 \pm 0.8) \times 10^{-5}$\\
 \hline
 CO$_{2}$ & $(1.3 \pm 0.7) \times 10^{-5}$  \\
 \hline
 CH$_{3}$OH & $(5.2 \pm 2.4) \times 10^{-6}$  \\
 \hline
  OCS & $(6.0 \pm 3.0) \times 10^{-8}$  \\
 \hline
 SO$_{2}$ & $(2.0 \pm 1.0) \times 10^{-6}$  \\
 \hline
 H$_{2}$S & $(8.0 \pm 4.0) \times 10^{-7}$  \\
 \hline
\end{tabular}
\caption{The abundances and uncertainties taken for the network with artificial sulphur constraints. For the first four species, the abundances were taken in their present form from \citep{Boogert}. \cite{Boogert} provided upper bounds for the listed sulphur-based species. For the analysis in this section, the abundances of the sulphur-based species were taken to be half the upper bound value. Their uncertainties were taken to be 50\%. }
\label{Extended_network_abundances}
\end{table}

\subsection{The Full Network}
Bayesian inference was performed for the full network with the new artificial constraints. This is Configuration 6. Despite adding these constraints elsewhere in the network, the maximum-posterior reaction rates of reactions 1-3 and 21-24 (none of which involve sulphur compounds) were found to be unchanged and the posteriors were largely similar. This fact strongly implies that the sulphur-based and non-sulphur-based reactions can be separated into sub-networks, whose reaction rates can be inferred independently, even when constraints on the sulphur-based products become available. 

\begin{figure*}[htp]
\centering
\includegraphics[width=\textwidth]{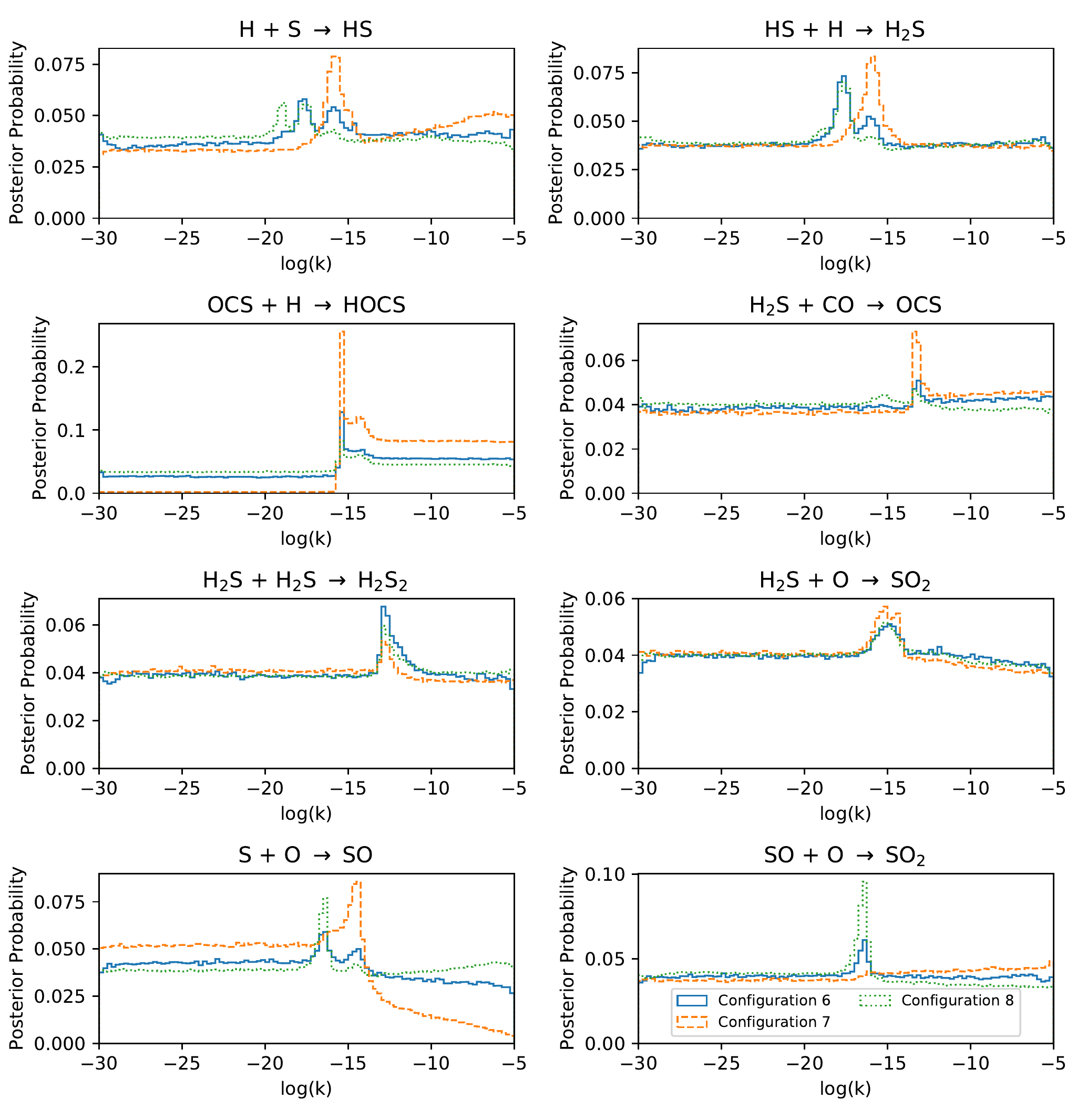}
\caption{Plots of the posterior probability distribution which deviate from uniformity for the expanded reaction network. We compare the posterior distributions of Configuration 6 with those of Configurations 7 and 8. We observe better agreement of the sulphur sub-network when we leave CO's abundance as a free parameter, which corresponds to Configuration 8.}
\label{new_sulphur_reaction rates}
\end{figure*}

We also identify eight new reactions for which the marginalised posterior probability distributions deviate from uniformity. These are all reactions that involve several of the molecules whose abundances have now been constrained. The posterior probability distributions are shown in Figure \ref{new_sulphur_reaction rates}.

\subsection{Including the CO constraint 1}\label{including_CO_constraint1}
In the following, we investigate the optimal way of splitting the full network into sulphur- and non-sulphur-based sub-networks. The two sub-networks compete over CO, one of the molecules with a constraint. We know from Section 5 that we can include the full CO constraint in the non-sulphur sub-network without significantly biasing the inferred reaction rates. To test if this is the case for the sulphur sub-network, we consider two cases, performing Bayesian inference on the sulphur sub-network with the full CO constraint (Configuration 7) and leaving the CO abundance as a free parameter (Configuration 8). The time taken for Configuration 6 is about 30 hours, whilst the runs for Configurations 7 and 8 took about 23 hours.

In Figure \ref{new_sulphur_reaction rates} we compare the rate posteriors for the artificially constrained sulphur-based reactions obtained with Configurations 6-8. For the eight new rate posteriors obtained in Figure \ref{new_sulphur_reaction rates}, we observe better agreement when the CO constraint is not included. The main explanation for this is that as several CO depletion reactions in Configurations 7 and 8 have been discarded, the other reactions in the network are required to produce more molecules that will react to deplete the CO abundance. This can be seen by the fact that the reaction rates for the successive hydrogenation of sulphur to produce H$_{2}$S are greater when the CO constraint is included. This is because H$_{2}$S reacts with CO to produce OCS. To deplete the excess CO, more H$_{2}$S must be produced.

It is interesting to note that one can apply the full CO constraint in the non-sulphur sub-network, but not in the sulphur sub-network. This is due to the relative sizes of the abundances of the constrained species. As shown in Table \ref{Extended_network_abundances}, the abundances of the sulphur-based molecules that are added are between 2 and 3 orders of magnitude less abundant than the constrained species in the other network. It should be noted that it is assumed that CO is already present to begin with and can only be consumed. No CO-formation reactions are present. As such, the contribution of the sulphur-network in depleting the CO is small compared to the non-sulphur network.

To get a better idea of the amount of CO that is used up by the sub-networks, we ran the forward models of the grain-code. By setting the reaction rates to zero, the only rates left were the freeze-out rates. These gave an idea of the total amount of CO available. This was found to be $4.0 \times 10^{-5}$. From Table \ref{first_abundance_table}, we know that the amount of CO that should be left is $(1.2 \pm 0.8) \times 10^{-5}$. Running the forward model of the sulphur-only network with CO as a free parameter, shows that the final CO abundance is about $(3.3 \pm 1.1) \times 10^{-5}$. This suggests that the non-sulphur network consumes four times as much CO as the sulphur-centric network does. By considering how the two sub-networks rely on the common constrained molecule, CO, we find that we can easily separate them.

Leaving the CO abundance unconstrained reduces the bias in the the maximum-posterior reaction rates, but this does not perfectly reproduce the full network's posterior. This is an issue, because it means that the variance of the full network is not preserved. An additional reaction of the form $X+CO \rightarrow XCO$ could potentially be used, just as was done in section \ref{variance_recovery}. However, unlike before this reaction will be replacing a reaction sub-network with constraints. This makes the problem more complicated than before, as one might need to consider how to combine the constraints to create a ``constraint" for XCO. One could simply give XCO an abundance equal to the sum of all the constraints that have been replaced. One might also need to provide several ``dummy" reaction chains of varying length to best recover the original posterior. Considerations of the architecture will be discussed in future work. 

\begin{figure*}[htp]
\includegraphics[width=\linewidth]{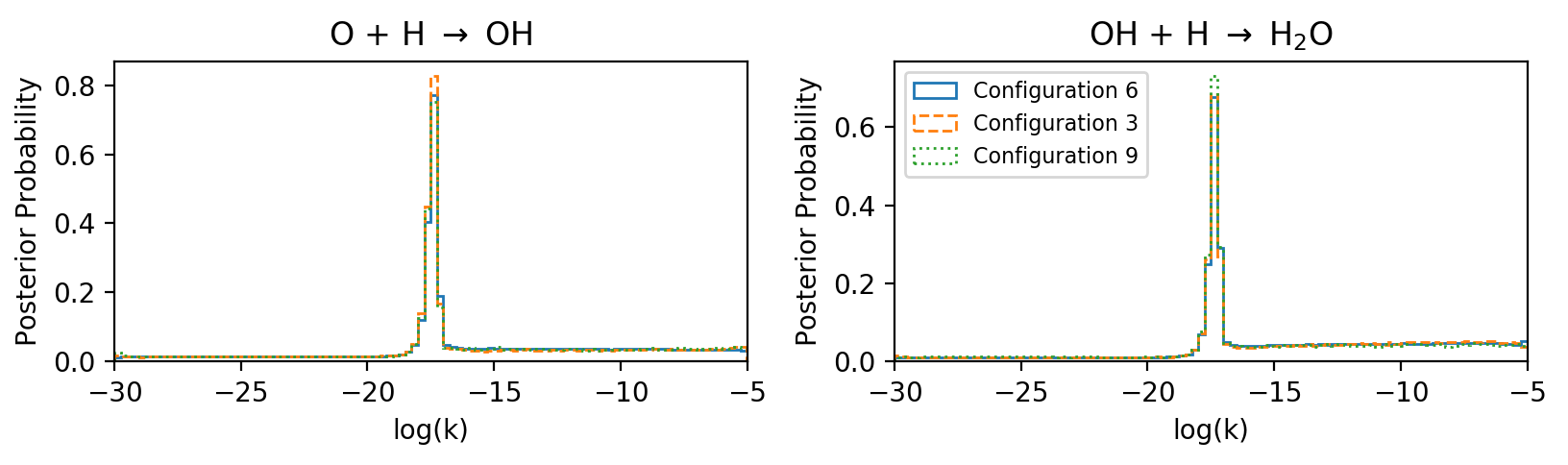}
\smallskip
\caption{The posterior distributions for Reactions 1 and 2 when the initial CO abundance is reduced by a factor of $10^{4}$. We compare the posterior distributions of Configuration 6 with those of Configurations 3 and 9. We observe the best agreement between Configurations 6 and 9, suggesting that for this sub-network, it is better to exclude the reduced CO constraint.}
\label{Non_sulphur_separation}
\end{figure*}

\subsection{Including the CO constraint 2}\label{includingCOconstraint2}
In the previous subsection, we argued that by virtue of the fact that the constraints in one sub-network were orders of magnitude greater than in the other sub-network, we could simply take the full CO constraint and use it in the former. Specifically, we want the CO constraint to be comparable to that of OCS, a molecule which depletes CO. However, we now consider the case where the constraints in each sub-network are comparable in nature. To do this, we artificially reduce the abundance of CO in the system. This is done by reducing the freeze-out rate by a factor of $10^{4}$. As only grain surface reactions are considered, this reduces the amount of CO available for grain-reactions by the same factor. 

Obviously changing the amount of CO present in the model has an effect on reactions obtained. With so little CO, it is impossible to match the abundances of methanol and carbon dioxide. As a result of this, the posteriors of these reaction chains are close to uniform. However, the point of these simulations is not to model the chemistry accurately. We want to know what we should do with the CO constraint in each sub-network. To aid understanding, we will consider the two sub-networks separately. We would like to note that even though the CO constraint has been reduced, the configuration for the full-network still corresponds to Configuration 6. 

\subsubsection{The non-sulphur sub-network}
Recall that the non-sulphur sub-network consists of reactions 1-3 and 21-24. The question is whether or not one wishes to include the reduced CO constraint to recover the posterior of the full-network (Configuration 6). Including the CO constraint gives Configuration 3 and excluding it gives Configuration 9. The posterior distributions are shown in Figure \ref{Non_sulphur_separation}. We observe that the only posteriors that deviate from non-uniformity are those for the reaction rates of Reactions 1 and 2. We observe that Configuration 9, which corresponds to excluding the reduced CO constraint, matches the posterior distribution of Configuration 6 the best for Reaction 1, not only in terms of the location of the maximum-posterior but also in terms of the posterior shape. Configuration 3 is found to recover the posterior better for Reaction 2.

We also observe that the maximum posterior reaction-rates for Reactions 1 and 2 match those obtained previously in the work, even though the CO abundance was greatly reduced. This adds support to the idea that Reactions 1 and 2 form their own sub-network and are ultimately independent.

\subsubsection{The sulphur sub-network}

\begin{figure*}[htp]
\centering
\includegraphics[width=\textwidth]{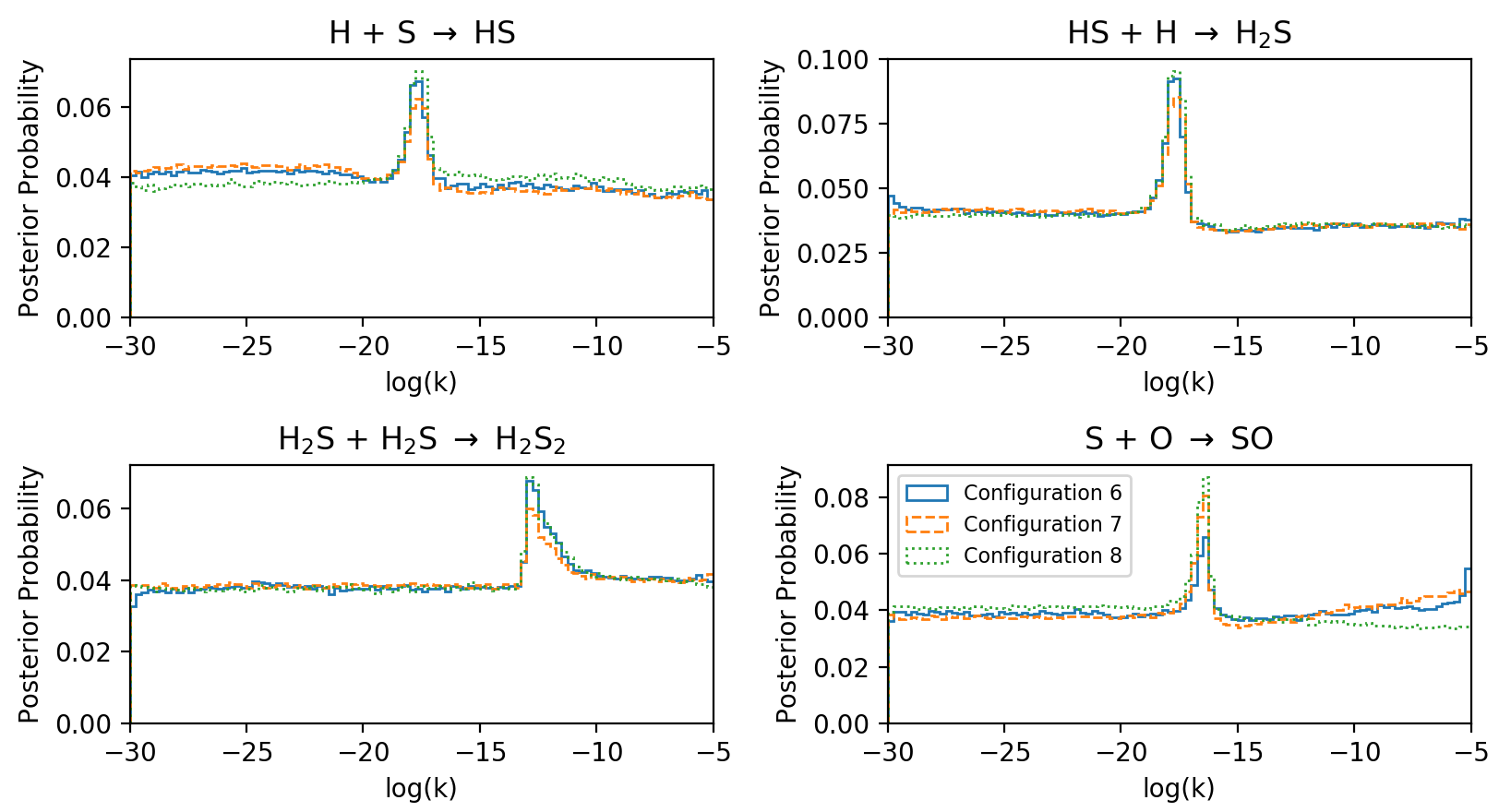}
\medskip
\caption{Plots of the obtained posteriors for Configurations 6,7 and 8 when the CO abundance is reduced by a factor of $10^{4}$. We observe that for this case, where the CO constraint is not several orders of magnitude greater than the abundances of the molecules in the sulphur sub-network, that it does not matter whether we include the CO abundance or not. Both cases allow us to recover the reaction rate with a very small bias.}
\label{Equal_CO_split}
\end{figure*}

Figure \ref{Equal_CO_split} shows the non-uniform posteriors of the reaction rates for the sulphur sub-network. We find that the maximum-posterior rates for these reactions do not change when we discard Reactions 1-3 and 21-24, regardless of whether we include CO’s new abundance constraint. As before, however, the precise forms of some of the posteriors are very different. This did not appear to be as much of an issue in Figure \ref{Non_sulphur_separation}, which might be related to the relative levels of uncertainty on the relevant species, as discussed in section \ref{network_reduction}.

\subsection{Comments on the Topology of the Sulphur Sub-Network}
We notice that adding the artificial constraints on sulphur-based species results in the rate of the reaction between H and OCS being constrained. This is interesting, as the abundance of the product, HOCS, is not constrained. Instead it is the penultimate molecule in the reaction chain, OCS, that is constrained. This suggests that it is not necessary for the end of a reaction chain to be constrained to constrain the reaction rates, as was observed with methanol in section 5.2. It seems that having a constraint on the penultimate molecule is sufficient. Constraining an earlier molecule would not do the trick, as was demonstrated when methanol's constraint was removed but CO's was kept. There is a notion of distance that needs to be considered. However, this would need to all be reconsidered for the case where there is more than one depletion mechanism for OCS. It is likely that having two depletion mechanisms, each of whose end product is unconstrained, would have a different effect, as there will be uncertainty about the branching ratio of each depletion route.


\section{Conclusions}

In this work, we have proposed new methods for performing Bayesian inference on chemical reaction networks that have very few constraints. We find that reducing the reaction network to just the reaction chains whose ends are constrained allows us to greatly reduce the computational expense. Despite the simplification, our most likely reaction rate values are mostly unchanged. We also find that we can separate chemical reaction networks into sub-networks, which can be analysed in parallel. We believe that such network reduction techniques will be of great use when looking at grain surface reaction networks, where there are few constraints on the molecules. However, it should be noted that the results of such a simplified chemical model can only provide a qualitative understanding of the chemistry.

We briefly summarise some general observations we have made that might prove useful in reducing reaction networks for Bayesian inference: 


\begin{itemize}
  \item Reducing the network reduces the computational expense of the inference process. The time taken for our inference runs scales roughly linearly with the number of reactions. However, the network connectivity is also likely to be a significant factor. This warrants further investigation.
  \item Reducing the network comes at the cost of artificially changing the variance of the posteriors. However, this variance can be partially recovered by adding a dummy reaction where the product is unconstrained, though there is the risk that the joint posterior distribution becomes more unrealistic. This needs to be investigated in further depth.
  \item When considering a reaction chain, it is important to include constraints for the final or penultimate molecule produced, as this ensures (for the case of a linear reaction chain) that the reaction rates of intermediate reactions are constrained. This might provide a general idea for future observations in terms of which molecules in the ices to look for. However, for more complex reaction networks, the intermediate reactions may play a more significant role.
  \item A network can be ``separated" into sub-networks. This is a potentially very useful tool, especially when looking at the grain surface chemistry of complex organic molecules, where the networks themselves are very large. For example, \cite{Garrod} provides a potential surface reaction network with around 200 reactions. In principle, this network could be split up into smaller sub-networks and Bayesian inference could then be performed on each sub-network in parallel. A potential general strategy would be to perform the network splitting at the point in the network with the highest network connectivity. For the network considered in this paper, this was the CO molecule. By making appropriate arguments about the relative magnitudes of the sub-networks and placing appropriate cuts in the networks, one could repeat the procedure as above. In order to decide what to do with a constrained molecule that is shared by the two sub-networks, one can make arguments about the relative abundances of the molecules in each sub-network. There are two cases to consider: 
   \begin{itemize}
     \item For the case where the shared constrained molecule has a significantly larger abundance than the molecules in one of the sub-networks, one can include its constraint in the higher-abundance sub-network.
     \item For the case where the shared constrained molecule is roughly the same as the abundances in either of the two sub-networks, one can choose to include it in either or both sub-networks.
   \end{itemize}
   For the case of a more interconnected networks with linear components (see Figure 11 in \cite{Linnartz} for an example of such a network), it makes sense to separate out the linear reaction chains first, as we found in section \ref{experimenting_with_linear_topology} that their topology is easy to understand intuitively. By removing these linear reaction chains, one could treat the interconnected sub-networks separately. Depending on their topology, one can employ the separation strategies discussed in this work.
\end{itemize}

Further work will need to focus on recovering the posterior distributions better. A quantitative approach is needed to better compare the posteriors inferred with full networks and sub-networks as well as explore how to use ``dummy reactions" of the form $X+CO \rightarrow XCO$ to recover the variance for the case where the reaction replaces a sub-network with constraints. 

Future work will also need to to look reaction networks with more complex geometries. The example considered here is fairly simple, with a relatively low degree of connectivity. As the complexity of the reaction networks considered increases, there will need to be more well-defined notions of how the position of a constraint influences the inference of related reaction rates. The guidelines we have presented are applicable to simple networks. Investigation of more interconnected networks is the focus of ongoing work. We aim to come up with a set of criteria to determine how to best separate more complex networks. 




\acknowledgments
The authors thank the referees for their constructive comments that greatly improved this work. J. Heyl is funded by an STFC studentship in Data-Intensive Science. This work was also supported by European Research Council (ERC) Advanced Grant MOPPEX 833460. S. Viti acknowledges support from the European Union’s Horizon 2020 research
and innovation programme under the Marie Skłodowska-Curie grant
agreement No 811312 for the project ``Astro-Chemical Origins” (ACO). S.M. Feeney is supported by the Royal Society. This work used computing equipment funded by the Research Capital Investment Fund (RCIF) provided by UKRI, and partially funded by the UCL Cosmoparticle Initiative.
\software{UCLCHEM \citep{UCLCHEM_release_paper}, PyMC3 \citep{pymc3}, F2PY \citep{F2Py}} 



\appendix
\section{Convergence}
Any MCMC chain needs to be checked for convergence. In the limit of an infinitely long chain, the sampled posterior distribution can be said to approximate the true posterior. However, when dealing with a finitely long chain, one needs to check that the posterior is not changing by much. We made use of two diagnostics to check. However, it should be emphasised that these diagnostics do not guarantee that the the chains are converged. Rather, if these checks fail, then we know the chain has not converged. Satisfying the conditions of the diagnostics simply lends credence to the hypothesis that the chains have converged.


\subsection{Geweke Diagnostic}

\begin{figure}[h]
\centering
\includegraphics[width=0.5\textwidth]{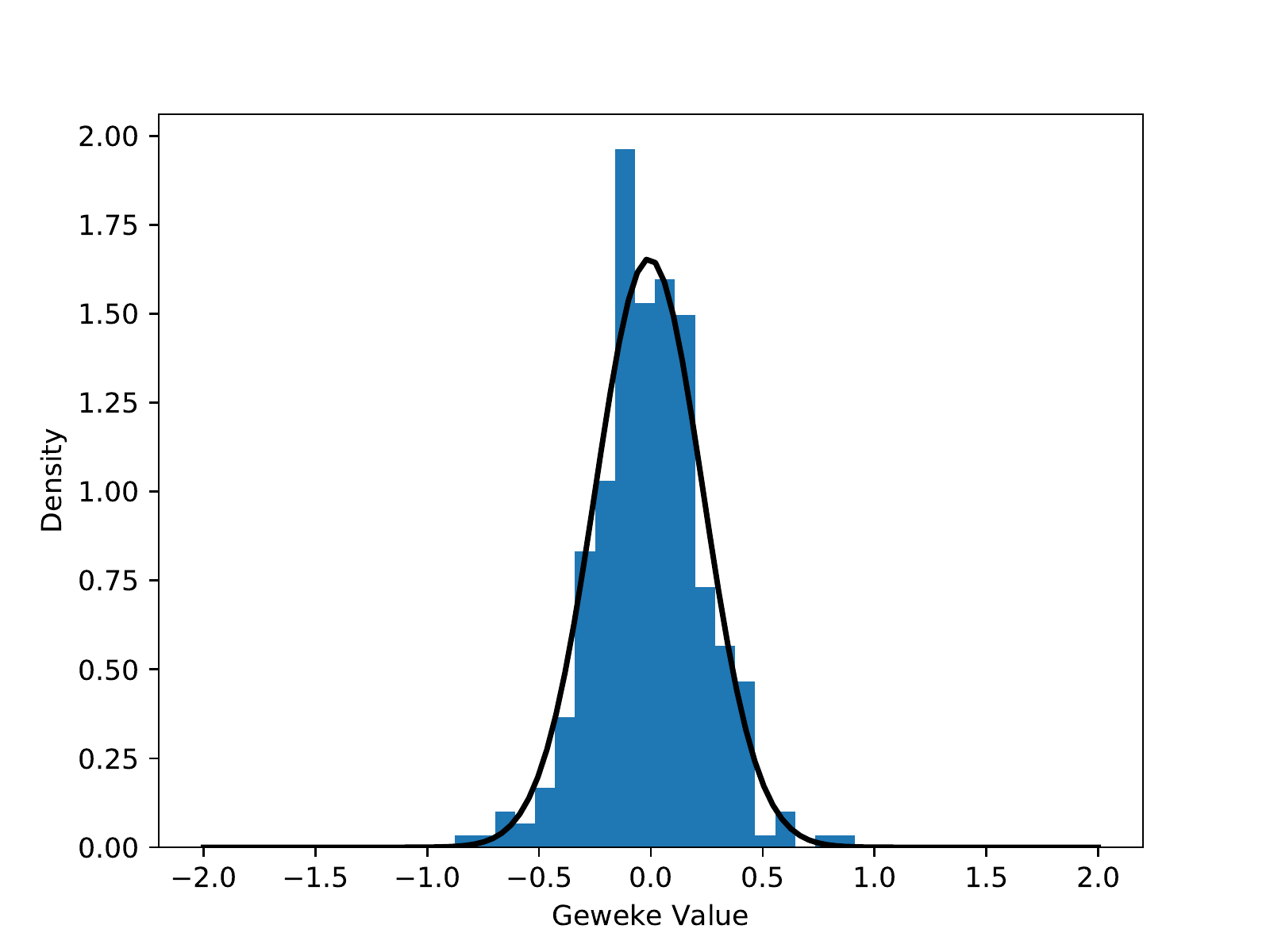}
\caption{A plot of the distribution of the Geweke diagnostic for the samples obtained in Configuration 6. We find that most of the points are within 2 z-scores of the mean. A normal distribution with zero mean is overlaid to show that the Geweke diagnostic is approaching a normal distribution.}
\label{geweke}
\end{figure}

The Geweke diagnostic calculates a z-score between two sections of a chain, typically the first 10\% and the final 50\% (\cite{Geweke, Roy}). The z-score is calculated by 
\begin{equation}
z = \frac{\overline{\theta_{a}} - \overline{\theta_{b}}}{\sqrt{\overline{\sigma_{a}^{2}}+\overline{\sigma_{b}^{2}}}},
\end{equation}
where the quantities with subscript $a$ refer to the first 10\% of the chain and the quantities with subscript $b$ refer to the final 50\%.
In the limit of the chain length going to infinity, the Geweke diagnostic is expected to follow a normal distribution (\cite{Cowles}). In Figure \ref{geweke}, we plot the distribution of the Geweke diagnostic for the chains of configuration 1 along with a normal distribution overlaid. We observe that the vast majority of points diagnostic stay within one standard deviation of the mean.

\subsection{Gelman-Rubin}
The Gelman-Rubin diagnostic provides a means of comparing the variance across all chains with the variance of the individual chains \cite{gelman1992, Hogg_2018}. There are several quantities of interest at play here. They are calculated for each scalar parameter of interest separately (\cite{gelman1992, Gelman2011InferenceFS}).

For $m$ chains, assumed to be of length $n$, the between-chain variance is defined as
\begin{equation}
B = \frac{n}{m-1} \sum_{i=1}^{m} \left(\widehat{\theta_{i}}-\widehat{\theta}\right)^{2},
\end{equation}
where $\widehat{\theta_{i}}$ is the estimator of the mean for chain $i$ and $\widehat{\theta}$ is the estimator of the mean of the sample. The latter is simply the average of all chain mean estimators. 

The within-chain variance, $W$, is defined as the average of the variances of all chains
\begin{equation}
W = \frac{1}{m} \sum_{i=1}^{m} \widehat{\sigma_{i}^{2}},
\end{equation}
where $\widehat{\sigma_{i}^{2}}$ is the estimator of the variance for chain $i$. 

The pooled variance estimate is defined as
\begin{equation}
\widehat{V} = \frac{n-1}{n}W + \frac{B}{n}. 
\end{equation}
The quantity of interest is 
\begin{equation}
\widehat{R} = \frac{\widehat{V}}{W},
\end{equation}
which is referred to as the potential scale reduction factor (PSRF). In the limit of infinitely long chains, $\widehat{R}$ tends to 1 from above. The closer $\widehat{R}$ is to 1, the better. In practice, a cut-off is used to determine convergence, typically 1.1 though these is some debate surrounding this (\cite{Vats, Roy}). The PSRF is related to the autocorrelation time that was used in H18, in that a value of $\widehat{R}$ that satisfies the criterion corresponds to a chain length greater than the autocorrelation time. There exists some debate as to whether to discard the first half of the chain when evaluating $\widehat{R}$ (see \cite{Roy} and \cite{Gelman2011InferenceFS} for two opposing views on the matter). Regardless of whether we do this, we find $\widehat{R} \leq 1.09$ for the reaction rates of the non-constrained reactions and $\widehat{R} \leq 1.03$ for the constrained reactions.
\section{Bayesian Sensitivity Analysis}\label{bayesian_sensitivity_analysis}
Bayesian parameter inference depends on two fundamental quantities: the likelihood and the prior. It is important to understand the relative information content of these two quantities to determine whether one's conclusions are driven by one's initial beliefs or the data at hand. In this work, we have used a log-uniform prior in the reaction rates, k, encoding our ignorance of their order of magnitude \citep{Jeffreys}. One of the key aspects when performing Bayesian inference is to determine if the posterior distributions are driven by the data or by the prior. This was considered in detail by \cite{Fischer} in the context of chemical kinetics, but has also been discussed with regards to climate models in \cite{Tomassini}. In this work, we motivated our choice of prior using chemical considerations and therefore used a log-uniform prior in the reaction rates \textbf{k} \citep{Chuang,Ioppolo}. \cite{Fischer} argues that if we are completely ignorant about \textbf{k}, then we are also ignorant in $f(\textbf{k})$, where $f$ is an arbitrary function in \textbf{k}. We could equivalently use a prior that is uniform in $f(\textbf{k})$. If the posterior distributions differ depending on the choice of prior, this means that the posteriors are ``prior-driven", as opposed to ``data-driven". 

\begin{figure}[h]
\centering
\includegraphics[width=\textwidth]{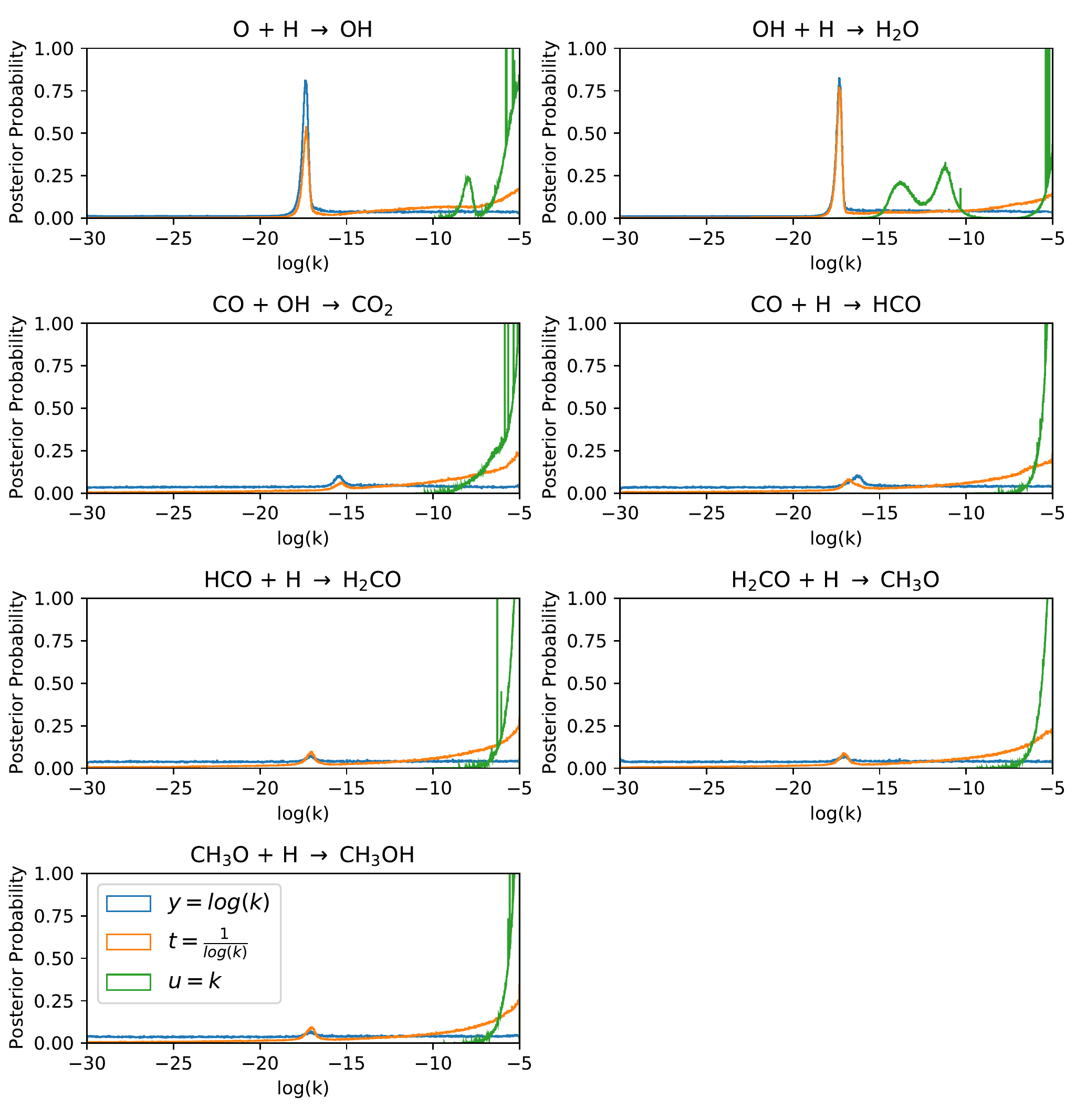}
\caption{Posterior probability distributions for Configuration 1 using three different priors. Alongside the uniform prior in $y=log(k)$, we also consider uniform priors in the variables $t=\frac{1}{log(k)}$ and $u=k$.}
\label{BSAconfig1}
\end{figure}

\begin{figure}[h]
\centering
\includegraphics[width=\textwidth]{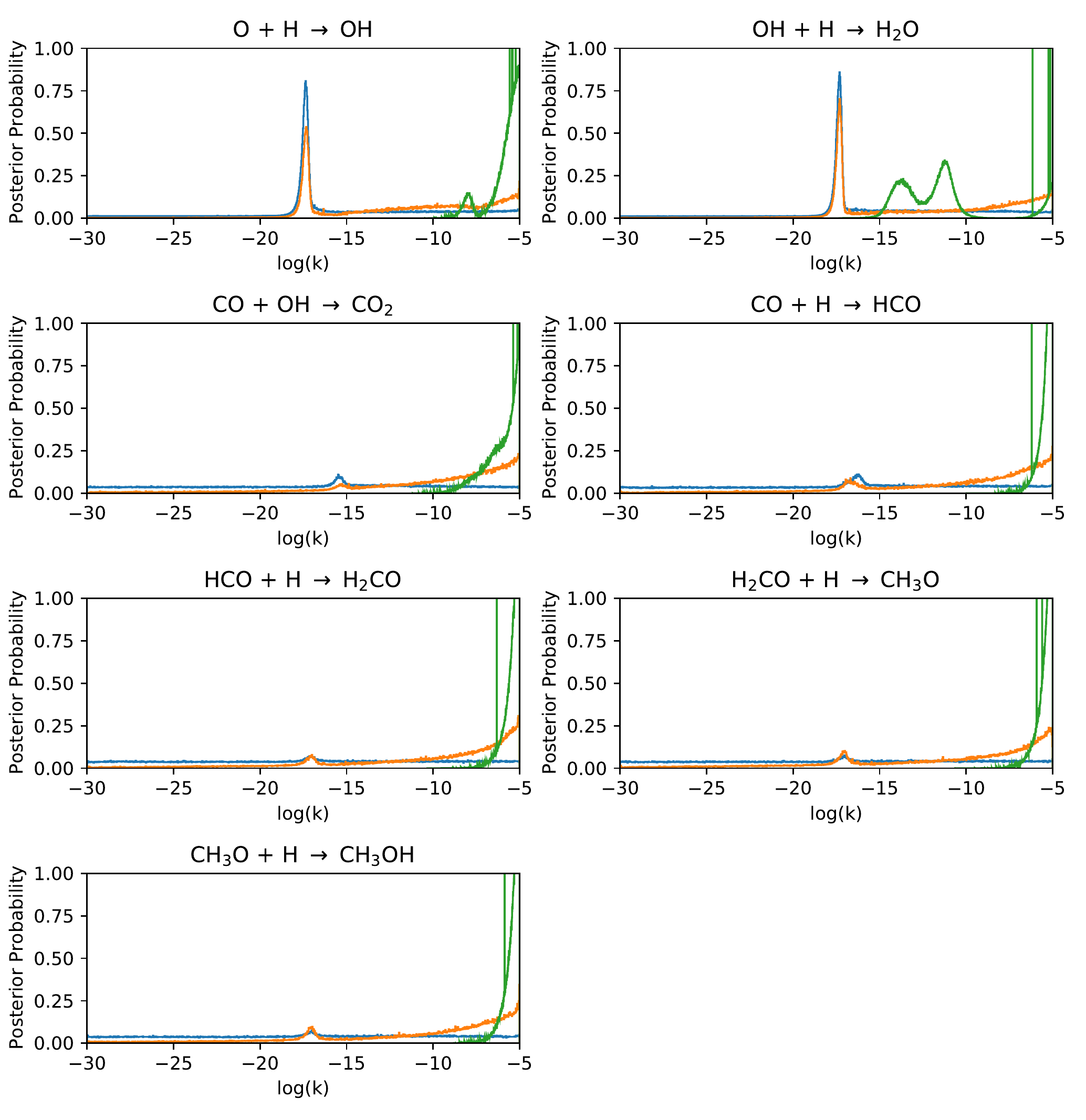}
\caption{Posterior probability distributions for Configuration 2 using three different priors. Alongside the uniform prior in $y=log(k)$, we also consider uniform priors in the variables $t=\frac{1}{log(k)}$ and $u=k$.}
\label{BSAconfig2}
\end{figure}

In this appendix, we repeat a portion of our analysis using two alternative priors in order to demonstrate the impact of different prior assumptions. Figure \ref{BSAconfig1} shows the posterior distributions for Configuration 1 using three different priors. Alongside the uniform prior in $y=log(k)$ that we have used throughout this paper, we also use uniform priors in $t=\frac{1}{log(k)}$ and $u=k$. We observe that the posteriors can differ significantly depending on the prior. However it is interesting to note that the posteriors for reactions 1 and 2 are relatively similar when uniform priors in $y$ and $t$ are used. As was discussed in section \ref{network_reduction}, this is due to the fact that these reactions are related to the formation of $H_{2}O$, the abundance of which is known to differ from zero at the $3\sigma$ level. This appears to suggest that the marginal probability distributions for Reactions 1 and 2 are more ``data-driven" than the posteriors for the other reactions. The other reactions are associated with the weaker abundance constraints. Their associated posterior distributions do not rule out much of the parameter space, so it is unsurprising that the posteriors are affected by the priors. We observe similar trends for the posterior distribution for Configuration 2, shown in Figure \ref{BSAconfig2}. We have cropped the ordinate at 1 in order to make sure the posteriors that came from using uniform priors in $y$ and $t$ are more visible.

\pagebreak

\pagebreak

\pagebreak


\bibliography{references}{}
\bibliographystyle{aasjournal}

\end{document}